\newcommand{\etal}{{\it et al.}}
\documentclass{PoS}

\title{\boldmath S-waves and the extraction of $\beta_s$}

\ShortTitle{S-waves}

\author{\speaker{Sheldon Stone}\\
        Department of Physics, Syracuse University\\
        Syracuse, N. Y., U. S. A., 13244\\
        E-mail: \email{stone@physics.syr.edu}}

\abstract{The $CP$ Violating asymmetry in $B_s$ mixing ($\beta_s$) is one of the most promising measurements where physics beyond the Standard Model could be revealed. As such, analyses need to be subjected to great scrutiny.
 The mode $B_s\to J/\psi \phi$ has been used, and the mode $B_s\to \phi \phi$ proposed for future measurements. These modes both have two vector particles in the final state and thus angular analyses must be used to disentangle the contributions from $CP+$ and $CP-$ configurations. The angular distributions, however, could be distorted by the presence of S-waves masquerading as low mass $K^+K^-$ pairs, that could result in erroneous values of $\beta_s$. The S-waves could well be the result of a final state formed from an $s$-quark $\overline{s}$-quark pair in a $0^+$ spin-parity state, such as the $f_0(980)$ meson.  Data driven and theoretical estimates of the  $B_s$ decay rate into the $CP+$ final state $ J/\psi f_0(980)$ are given, when $f_0\to\pi^+\pi^-$.  The S-wave contribution  in $J/\psi\phi$ should be taken into account when determining $\beta_s$ by including a $K^+K^-$ S-wave amplitude in the fit. This may change the central value of current results and will also increase the statistical uncertainty. Importantly, the $J/\psi f_0(980)$ mode has been suggested as an alternative channel for measuring $\beta_s$.  }

\FullConference{Flavor Physics and CP Violation - FPCP 2010\\
		May 25-29, 2010\\
		Turin, Italy}

\begin{document}
\section{Introduction}
Measurements of Charge-Parity ($CP$) violation in the $B_s$ meson system are sensitive to the presence of heavy, as yet undiscovered, particles.  The $CP$ violating angle $-2\beta_s$, the so called ``phase of $B_s-\overline{B}_s$ mixing" is a particularly important place to look for physics beyond the Standard Model, since the expected asymmetry is very small,  $2\beta_s=0.036\pm 0.002$ \cite{Charles}, thus allowing the effects of any new physics to be more easily observed.
Both CDF \cite{CDF-chi} and D0 \cite{D0-chi} have investigated $-2\beta_s$ using $B_s\to J/\psi\phi$ decays. Central values have been found far from the expected Standard Model values, but the errors are large and the results are not statistically significant.

Since the final state consists of two spin-1 particles, it is not a $CP$-eigenstate, yet it is well known that $CP$ violation can be measured using angular analyses \cite{transversity}. Except for one very recent analysis \cite{CDF-FPCP}, previous determinations have ignored the possibility of an S-wave $K^+K^-$ system in the region of the $\phi$. Not accounting for the S-wave can bias the result, and the resulting quoted error is smaller than if the S-wave components are allowed in the fit.

\section{Evidence for S-waves}
\begin{figure}
\centering
\includegraphics[height=3.5in]{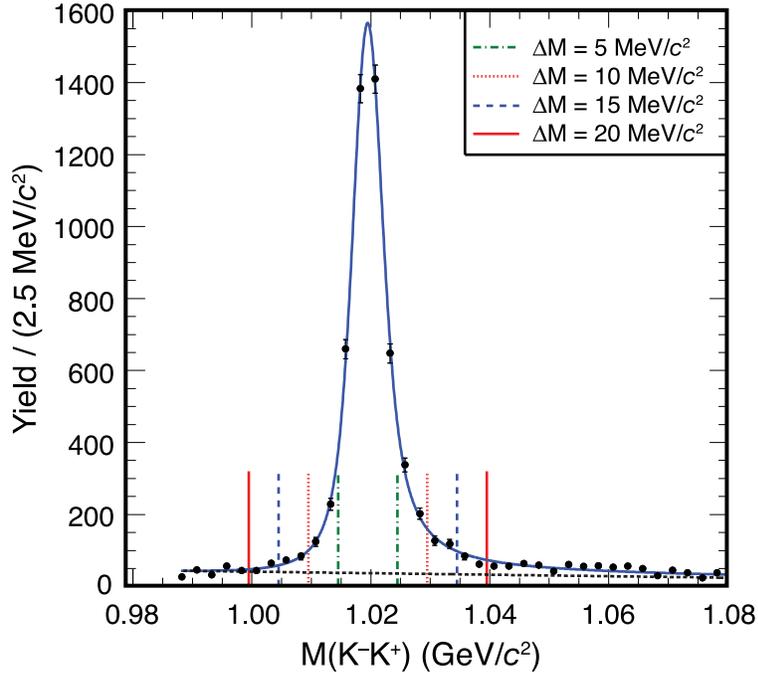}
\caption{Dalitz plot projections for $K^+K^-$ invariant mass in $D_s^+\to K^+ K^-\pi^+$ from the CLEO Collaboration \cite{CLEOabs}. The signal is extracted individually in each mass bin, thus there is no background. The data are fit with a Breit-Wigner signal function for the $\phi$ convoluted with a Gaussian for detector resolution and linear representation of an S-wave component (dashed line). The solid curve shows the sum. (Only the data is ascribed to CLEO, the fits have been added.)}
\label{CLEOmKK}       
\end{figure}

In fact, there is a great deal of evidence for S-waves in many decays where vectors are dominant.  Consider the reaction $D_s^+\to K^+K^-\pi^+$. This mode has long been known to have large $\phi\pi^+$ and $K^*K$ components \cite{PDG}.
CLEO has looked explicitly at the low mass $K^+K^-$ region in the $K^+ K^-\pi^+$ Dalitz plot \cite{CLEOabs}. Fig.~\ref{CLEOmKK} shows the $K^+K^-$ mass projection in the region near 1 GeV. The signal is extracted by fitting the $D_s$ yield in each bin of $K^+K^-$, so no background remains in the plot. There is an additional component of signal beneath the $\phi$. To estimate the size of this component,  the CLEO data  are fit to a Breit-Wigner to describe the $\phi$, convoluted with a Gaussian for detector resolution, and in addition a linear component that we take as an S-wave based on Dalitz plot studies. The fraction of S-wave depends on the mass interval considered. For $\pm$10 MeV around the $\phi$ mass there is a 6.3\% S-wave contribution, which rises to 8.9\% for a $\pm$15 MeV interval. (Note that these fractions depend on the experimental resolution.)

If the S-wave here is the  $0^+$ $f_0$(980) state, then we should see an $f_0$ signal peak in the $D_s^+\to\pi^+\pi^+\pi^-$ final state, since
the $f_0$(980) decays into $\pi^+\pi^-$ as well as $K^+K^-$.   A BaBar Dalitz plot analysis shows a large $f_0$ signal \cite{Babar-Dalitz}.

\begin{figure}
\centering
\includegraphics[height=2.9in]{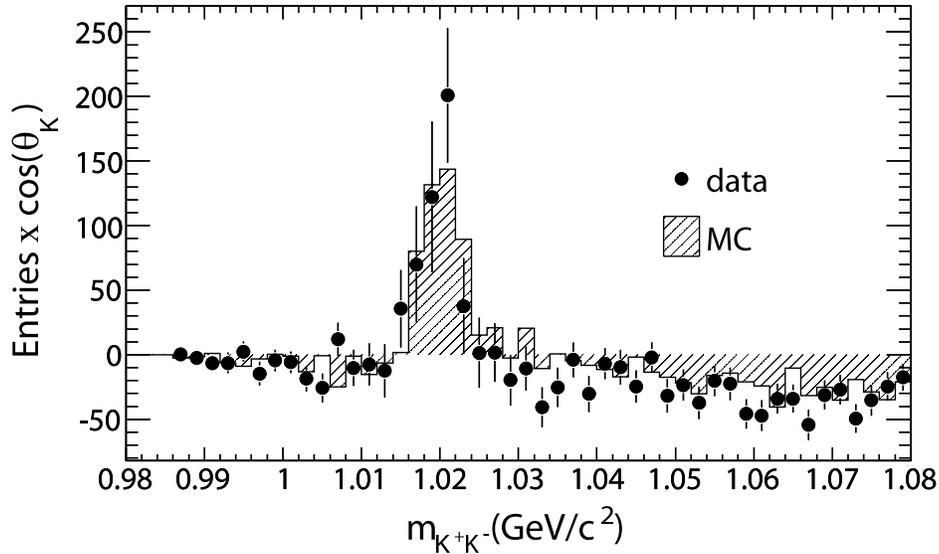}
\caption{$K^+K^-$ invariant mass distribution weighted by the measured value of the cosine of the decay angle from BaBar \cite{Babarsemi} in the channel $D_s^+\to K^+K^-e^+\nu$. }
\label{KK}       
\end{figure}

S-wave effects have also been observed in semileptonic charm decays.   FOCUS observed an interfering S-wave amplitude in the $D^+\to K^+\pi^-\mu^+\nu$ channel with a rate fraction of (2.7$\pm$0.4)\% with respect to the P-wave $K^*$ in the $K^-\pi^+$ invariant mass region between 0.8--1.0 GeV \cite{FOCUS}. BaBar measured an S-wave fraction in $D_s^+\to K^+K^-e^+\nu$ decays of  $(0.22^{+0.12}_{-0.08})$\%  for $K^+K^-$ invariant masses between 1.01--1.03 GeV \cite{Babarsemi}.  The $K^+K^-$ invariant mass spectrum when weighted by the $K^+$ decay angle shows a clear distortion due to the interference (Fig.~\ref{KK}).

\begin{figure}
\centering
\includegraphics[height=2.in]{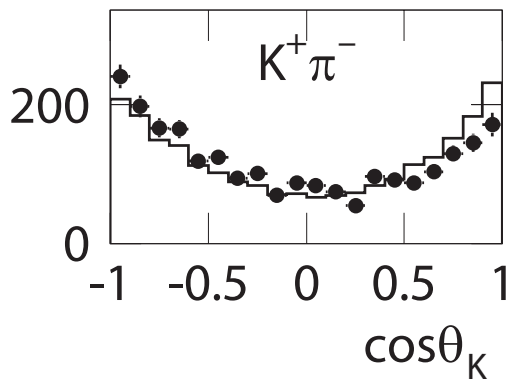}
\caption{The cosine of the decay angle of the kaon from the $\overline{K}^*$ decay in $B^0\to J/\psi \overline{K}^{*0}$ decay. The histogram is Monte Carlo without S-P wave interference, and the points the BaBar data \cite{Babar-psiKstar}. }
\label{angles-ch2}       
\end{figure}

The analogous channel to $J/\psi\phi$ in ${B}^0$ decay $J/\psi \overline{K}^{*0}$ is well known to have an S-wave $K\pi$ component in the $K^*$ mass region. This interference, in fact, has been used by BaBar to measure $\cos(2\beta)$ and thus remove an ambiguity in the value of $\beta$ from the $\sin(2\beta)$ measurement. The S-wave component in the region of the $K^*$ is measured as $(7.3\pm 1.8)$\% of the P-wave for $0.8<m(K\pi)<1.0$ GeV \cite{Babar-psiKstar}.
BaBar uses this interference to remove ambiguities in the measurement of $\cos(2\beta)$, where $\beta$ is the $CP$ violating angle measured in $B^0\to J/\psi K_S$ decays, for example. Visual evidence of the S-wave can be seen in Fig.~\ref{angles-ch2} where there is an obvious asymmetry of the decay angle distribution of the kaon in the $\overline{K}^{*0}$ rest-frame.

\begin{figure}[htb]
  \begin{center}
\includegraphics[width=4.5in]{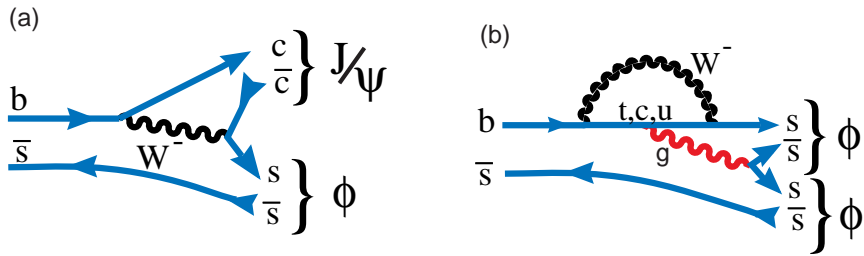}
   \end{center}
\vskip -0.15in
\caption{Decay diagrams (a) $B_s\to J/\psi \phi$, and (b) $B_s\to \phi\phi$.}
  \label{psi_ss}
\end{figure}

Perhaps it may be hoped that the S-wave $K^+K^-$ under the $\phi$ in $J/\psi\phi$ is smaller due to the relatively narrow width  ($\Gamma$) of the $\phi$ (4.3 MeV) compared to the $K^*$ (51 MeV), but even so, the question is how much does the presence of the S-wave $amplitude$ affect the extraction of $\beta_s$? Similar considerations apply to the measurement of $CP$ violation in the process $B_s\to \phi\phi$. Here the problem is exacerbated by the presence of two $\phi$'s in the final state. The decay diagrams for both of these processes are shown in Fig.~\ref{psi_ss}. In both cases the $s\overline{s}$ forms a $\phi$. Other manifestations of $s\overline{s}$ quarks are the $\eta$, $\eta'$ and $f_0$(980) mesons. The first two are pseudoscalars, while the last is a scalar.

The formalism for the time dependent $B_s$ and $\overline{B}_s$ decay rates as a function
of the decay angular distributions is given in Ref.~\cite{Form}. Addition of the S-wave amplitudes
was done by Xie \etal~\cite{Xie}. The number of terms to consider increases from 6 to 10.  Another approach has been given by Azfar \etal~\cite{Azfar}. Adding the S-wave terms in the fit can only increase the experimental error. The size of the effect depends on many factors including the magnitude and phase of the S-wave amplitude, $\beta_s$, values of the strong phases, detector acceptances, biases, etc..

While S-waves are a nuisance in analyzing the $J/\psi\phi$ final state, they can also be used beneficially. When the $f_0$ materializes as a $\pi^+\pi^-$ there cannot be an iso-vector $\rho$ state as $s\overline{s}$ pair is isoscalar. Near the $\phi$ mass the $f_0(980)$ can materialize as a $\pi^+\pi^-$ pair in the $0^+$ state and this $J/\psi f_0$ state is useful for $\beta_s$ measurements \cite{Stone-Zhang}.
The final state  is a $CP+$ eigenstate, thus no angular analysis is necessary! Note, that the modes $J/\psi \eta$ and $J/\psi \eta'$ can also be used, but they involve photons in the decay and thus have lower efficiency at hadron colliders.

Predictions of the ratio
\begin{equation}
R_{f_0/\phi}\equiv\frac{\Gamma(B_s^0\to J/\psi f_0,~f_0\to \pi^+\pi^-)}{\Gamma(B_s^0\to J/\psi \phi,~\phi\to K^+K^-)}
\end{equation}
have been given based on $D_s$ decay data, and purely from theory.  Stone and Zhang using  $D_s^+\to f_0 \pi^+$ decays where $f_0$ was detected in both $K^+K^-$ and $\pi^+\pi^-$ modes predicted $R_{f_0/\phi}\approx 20$\% \cite{Stone-Zhang}.  CLEO made an estimate of
$R_{f_0/\phi}=(42\pm 11)$\% based on the ratio of the branching fractions for $D_s^+\to f_0 e^+\nu$ to $D_s^+\to \phi e^+\nu$ at $q^2=0$ where the mass difference between the $D_s$ and the final state hadron is largest, which best approximates $B_s\to J/\psi$ decays \cite{CLEO-f0semi}.

Theoretical predictions for $R_{f_0/\phi}$ are difficult, however there are a few heroic attempts. Colangelo, De Fazio and Wang
use Light Cone Sum Rules to make two predictions \cite{CDW}. For the first they use their calculations of  ${\cal{B}}(B_s\to J/\psi f_0)$ which are
$(3.1\pm 2.4)\times 10^{-4}$ at leading order (LO) and  $(5.3\pm 3.9)\times 10^{-4}$ at non-leading order  (NLO), combined with the
measured ${\cal{B}}(B_s\to J/\psi\phi)=(1.3\pm 0.4)\times 10^{-3}$ to predict  $R_{f_0/\phi}$=24\%  (LO) and $R_{f_0/\phi}$=41\% (NLO).
Secondly, they use the form-factor calculation of Ball and Zwicky  \cite{BZ} to predict $R^L_{f_0/\phi}$=13\%  (LO) and $R^L_{f_0/\phi}$=22\% (NLO), where $R^L$ refers to only longitudinal $\phi$ production, so since transverse $\phi$ production is about 46\% this lowers their $R_{f_0/\phi}$ predictions for the second case by almost a factor of two \cite{PDG}.  The experimental predictions above for $R_{f_0/\phi}$ based on $D_s$ decays are also enhanced by normalizing to $\phi$ final states that are mostly longitudinal. Thus they should be lowered.

In a later paper using QCD factorization Colangelo, De Fazio and Wang \cite{CDW2}  predict  ${\cal{B}}(B_s\to J/\psi f_0)=(4.7\pm 1.9)\times 10^{-4}$ using CDSS form-factors \cite{CDSS}, and $(2.0\pm 0.8)\times 10^{-4}$ using Ball and Zwicky \cite{BZ} . These predictions are somewhat smaller than those given above, but still have $R_{f_0/\phi}$ as 36\% or 20\%.
Within the framework of QCD factorization O. Leitner \etal~\cite{Leitner} give a range of predictions for  $R_{f_0/\phi}$ that are in the 30-50\% range. Thus predictions for  $R_{f_0/\phi}$ have a rather wide range from 7-50\%.

The only reported experimental search for $J/\psi f_0;~f_0\to\pi^+\pi^-$ was done by BELLE using 23.6 fb$^{-1}$ of data taken on the $\Upsilon(5S)$ resonance, about 1/5 of their total accumulated data sample. They find  $R_{f_0/\phi}<27.5$\% at 90\% c.l. \cite{Louvot}. Their data however show a hint of signal with a central value about half of the upper limit. It will be quite interesting to see which experiment finds the signal first.  CDF has recently put the S-wave amplitudes in their fits for $\beta_s$ in the $J/\psi\phi$ channel. They find that the fitted fraction of $K^+K^-$ S-wave in the signal
region is $<6.7$\% at 95\% c.l. \cite{CDF-FPCP}. They do not however, report any result for a direct search using the $f_0\to\pi^+\pi^-$ channel.

In conclusion, S-waves are ubiquitous, they appear whenever looked for and must be taken into account in $B_s\to J/\psi\phi$ measurements of amplitudes, phases, and $CP$ violation. Kudos to CDF for including S-waves in their most recent fits.
In addition it appears to be wise to add S-wave amplitudes in the analysis of $B\to K^*\mu^+\mu^-$ and surely in $B_s\to\phi\phi$.
Furthermore, especially since angular analysis is not required, $B_s\to J/\psi f_0$ may be a useful mode to add to the statistical precision on the measurement of $-2\beta_s$ and will provide a useful systematic check \cite{SZLHCb}.

\section*{Acknowledgments}
I thank the U. S. National Science Foundation for support and L. Zhang for
his collaboration on this work.

\end{document}